# Structured Spreadsheet Modeling and Implementation

Paul Mireault
Honorary professor, HEC Montréal
Montréal, Canada
Paul.Mireault@hec.ca

*Abstract* — Developing an error-free spreadsheet has been a problem since the beginning of end-user computing. In this paper, we present a methodology that separates the modeling from the implementation. Using proven techniques from Information Systems and Software Engineering, we develop strict, but simple, rules governing the implementation from the model. The resulting spreadsheet should be easier to understand, audit and maintain.

*Index Terms* — Spreadsheet modeling, Spreadsheet implementation, structured modeling.

## I. Introduction

*How have you learned Excel?* The majority of Excel users we asked in MBA and Executive training courses answer that they have learned it simply by using it, sometimes with the help of a colleague who had shown them a few *tricks*. A few had attended Excel classes, varying from a few hours to a couple of days. However, those classes usually show ***what*** Excel can do, and not ***how*** to use it in a business context. It is like learning a language by reading a dictionary: you may learn the meaning of the words (the *what*), but not the syntax (the *how*).

The same can be said about Excel books: they will explain all the features, menus and functions offered by the program. Some books will show how to use Excel in specific contexts, like Finance, Accounting, Marketing and Human Resources, but they present a limited number of spreadsheet templates that the reader is expected to modify in order to adapt them to his or her needs.

Excel presents the illusion of simplicity because of its *free-form* structure or, should we say, its lack of structure. This lack of structure may please the novice users, but as they become more experienced, they often impose upon themselves some form of structure. For example, they may indicate input cells with a particular colour, format its descriptive label (i.e. the text, usually on the left, that describes the cell's content) or even separate the data from the formulas in different areas of the spreadsheet. This very need for self-imposed structure is an indication that the free-form structure has its shortcomings.

Self-imposition of rules is not new. Many industries have adopted *codes* or *norms*. For example, think of what would happen if electricians did not have a norm for the color of wires: maintenance would be problematic, if not outright dangerous. Following norms makes us incur additional costs: extra inventory (i.e. spools of different coloured wires instead of all black) and other resources. Yet, nobody disputes the benefits of those norms; they let another electrician continue the work of the original installer, sometimes many years later.

IT departments often impose norms on their programmers such as variable naming conventions and indentation rules, to name a couple. If these norms weren't imposed, then maintenance would be problematic; the maintenance programmer would have to spend hours trying to understand both the original programmer's intent as well as the work of previous maintenance programmers.

Programmers, electricians, plumbers and engineers have all had training in which the importance of norms has been taught. None of the Excel users in our courses had.

In this article, we present a spreadsheet development methodology named *Structured Spreadsheet Modeling and Implementation* (SSMI). This methodology was taught in MBA, Executive training and undergraduate business courses. In the next section, we explain the underlying Information Systems concepts of the methodology. Then, we illustrate the methodology with an example.

## II. Modeling

Past research on spreadsheet design has focused on the physical design of the spreadsheet itself.

The analogy with computer systems is like talking about the design of a program in terms of declaring all the variables at the beginning, indenting loops and if-then-else structures or naming variables with descriptive names using a mix of lower and upper case letters. However, the core of a computer program is its algorithm, not how it looks. The same with spreadsheets: they are a model of a real-world system. All the formulas in the spreadsheet are the model, and very little has been said about the model's design.

In this article, we make a distinction between the design of the spreadsheet, which we will call the *spreadsheet implementation*, and the design of the model, which we will simply call *spreadsheet modeling*. We will use concepts from the *Information Systems* (IS) literature to explain a modeling methodology for spreadsheet developers.

A. Developing an Information System

Figure 1 illustrates a simplified view of information systems development where three models are used in succession. Even though the process appears to be sequential, there are feedback loops allowing changes in previous steps.

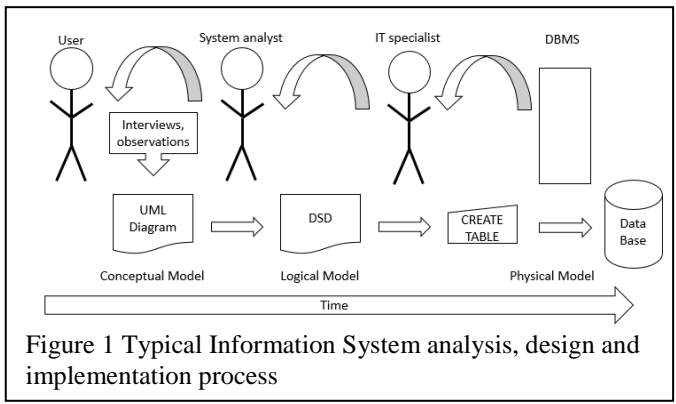

Figure 1 Typical Information System analysis, design and implementation process

The final model is called the *physical model* and is in fact the implementation. In a relational database system, it is the actual data tables created with the SQL statement CREATE TABLE (see Figure 2) and managed by the Database Management System (DBMS) like Oracle, Microsoft SQL Server or IBM DB2.

Before the creation of the tables, the system analysts produce the *logical model*, represented with a Data Structure Diagram (DSD) (see Figure 3), which is a high-level description of the data tables showing all their fields and specifying the relationships between the tables through the primary keys (PK) and the foreign keys (FK). The DSD is the tool with which the system analyst communicates with the IT specialists who will implement the database. It contains technical information that is not always understood by the system user like, for example, *primary key* and *foreign key*. While these concepts are essential for a proper implementation, it is not necessary that the user understands them.

Since the system is developed for the user, there must be a way for him to convey his needs and requirements down the chain. This is done through the *conceptual model*. In a database system, the conceptual model can be represented with a Unified Modeling Language Diagram (UML). An important characteristic of the conceptual model is that it is not dependant on any technology (Oracle, Microsoft or IBM) and it does not

```
CREATE TABLE CRUISE(
    CRUISECODE      INTEGER NOT NULL,
    CRUISENAME      VARCHAR2(100) NOT NULL,
    SHIPNO          INTEGER NOT NULL,
    DEPARTURE_PORT  VARCHAR2(100) NOT NULL,
    ARRIVAL_PORT    VARCHAR2(100) NOT NULL,
    CONSTRAINT CRUISE_PK
        PRIMARY KEY(CRUISECODE),
    CONSTRAINT CRUISE_FK_SHIP
        FOREIGN KEY(SHIPNO) REFERENCES SHIP(SHIPNO));

CREATE TABLE STOP(
    CRUISECODE      INTEGER NOT NULL,
    STOPNO          INTEGER NOT NULL,
    PORT            VARCHAR2(100) NOT NULL,
    ARRIVAL_DAY     INTEGER,
    ARRIVAL_HOUR    INTEGER,
    DEPARTURE_DAY   INTEGER,
    DEPARTURE_HOUR  INTEGER,
    CONSTRAINT STOP_PK
        PRIMARY KEY (CRUISECODE,STOPNO),
    CONSTRAINT STOP_FK_CRUISE
        FOREIGN KEY(CRUISECODE) REFERENCES
CRUISE(CRUISECODE));
```
Figure 2 Physical Model: SQL statements

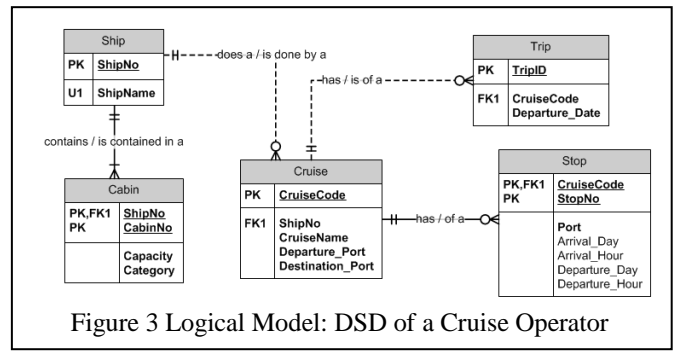

Figure 3 Logical Model: DSD of a Cruise Operator

use technical concepts, rather, it uses the user's vocabulary. The UML Diagram is usually designed by the system analyst, following interviews with the users and observation of the work processes they perform. In Figure 4 we can understand that a ship has cabins, that a cabin is contained in a ship and that a cruise has stops. A cruise has a set of stops and it has trips that sail on specific dates and are sold to clients.

B. Developing a Spreadsheet

With spreadsheets, there is often no system analyst or IT specialist present, as illustrated in Figure 5. The user is usually the developer and builds his spreadsheet by alternating between the creative activities like thinking about formulas and variables, and the mechanical activities, like typing formulas, pointing to cells, copying them, testing them and so on. As the spreadsheet grows in size, these mechanical operations take longer to perform and are a source of errors.

This is an unreliable development process because spreadsheet developers, even though they are their domain's specialists, are not trained in development techniques like IT specialists. In fact, they are part-time developers because developing spreadsheets usually only represents a small part of their job.

Furthermore, since users don't have formal training, it is left to themselves to build the spreadsheet as they see fit. Some organizations have developed standards for spreadsheet applications (see FAST[1]), but what they call a *model* is what we call the Physical Model.

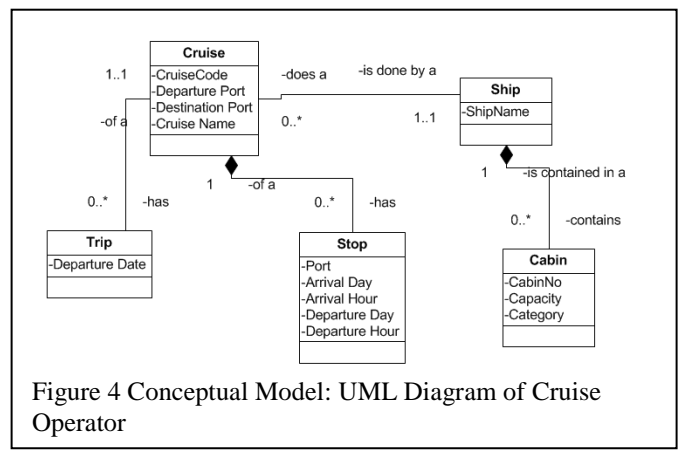

Figure 4 Conceptual Model: UML Diagram of Cruise Operator

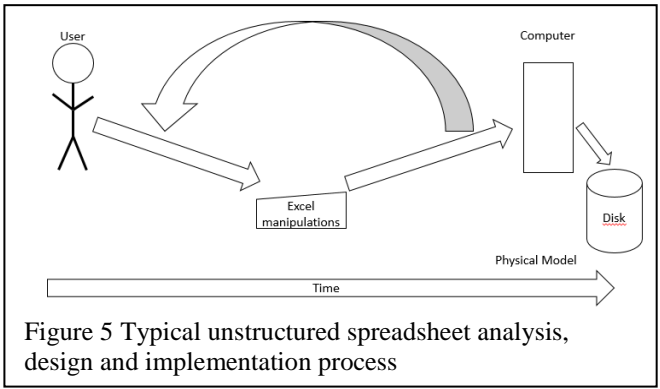

Figure 5 Typical unstructured spreadsheet analysis, design and implementation process

Because of this free-for-all approach, researchers have studied ways to audit or understand a spreadsheet's logic (Clermont[2], Hermans [3], Igarashi [4]). Unfortunately, this after-the-fact approach to understanding the spreadsheet developer's logic is difficult and time consuming.

Other research provided guidelines for the spreadsheet developer. Alexander [5] gives guidelines for the Physical Model.

### III. Structured Methodology

The structured methodology is based on the same concepts presented in Figure 1: a separation of tasks to produce three models. The process is illustrated in Figure 6.

The conceptual model is called a Formula Diagram (FD), the logical model is the Formula List (FL) and the physical model is the actual spreadsheet. Since the Formula Diagram and the Formula List have a one to one correspondence and require more domain knowledge than spreadsheet knowledge they are produced at the same time and by the same person. The Formula Diagram was inspired by Bodily's *Influence Diagram* [6]. Ronen's [7] *Spreadsheet Flow Diagrams* are a Conceptual Model, but its implementation does not follow directly from the diagrams.

We will illustrate the process by building a spreadsheet for a small North-American car rental company. A car has a daily cost and an additional cost if the client goes over the total distance allowance. The total distance allowance is the product of the number of days and the daily distance allowance.

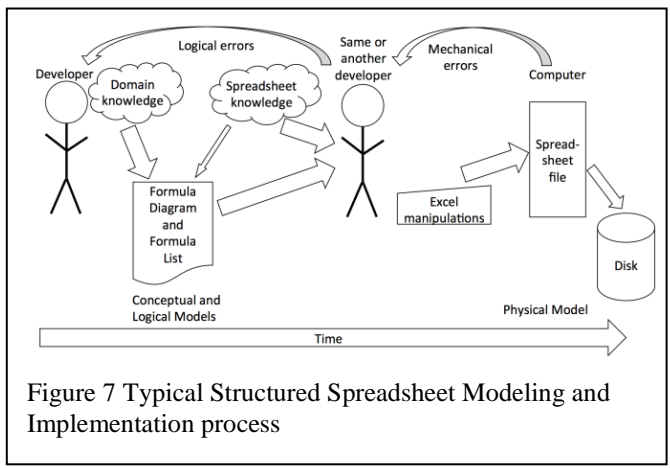

Figure 7 Typical Structured Spreadsheet Modeling and Implementation process

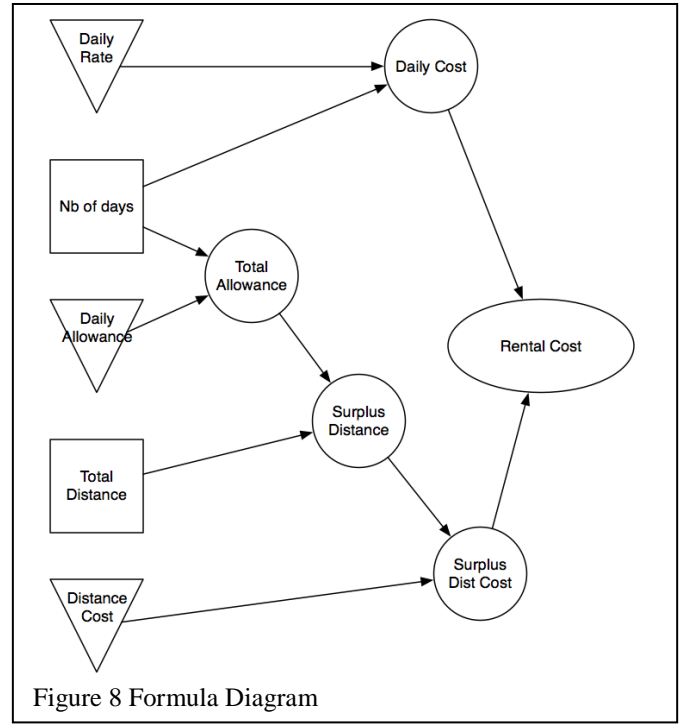

Figure 8 Formula Diagram

A. Modeling

We will start by developing the model without any information regarding the interface.

The Formula Diagram corresponding to our car rental company is illustrated in Figure 7.

The symbols have the following signification:
- A triangle represents a constant that is not changed often. We will put it in the Parameters sheet during the implementation.

| Variable | Type | Formula / initial value |
|---|---|---|
| Daily Rate | Input | 58$ |
| Nb Days | Input, Interface | 12 |
| Daily Allowance | Input | 100 |
| Total Distance | Input, Interface | 1452 |
| Distance Cost | Input | 0,36$ |
| Daily Cost | Intermediate | = Nb Days * Daily Rate |
| Total Allowance | Intermediate | = Nb Days * Daily Distance |
| Surplus Distance | Intermediate | = IF(Total Distance > Total Allowance; Total Distance – Total Allowance; 0) |
| Surplus Dist Cost | Intermediate | = Surplus Distance * Distance Cost |
| Rental Cost | Intermediate, Interface | = Daily Cost + Surplus Dist Cost |

Figure 6 Formula List

- A rectangle represents a constant that is changed during the regular use of the spreadsheet. It will be put in the Interface sheet.
- A circle is an intermediate variable that is defined by a formula using the variables sending arrows to it. It will appear in the Model sheet.
- An ellipse represents an intermediate variable that has a particular importance to the spreadsheet user. Its definition will be in the Model sheet, like all the intermediate variables, and there will be a reference to its value in the interface sheet.

As we develop the Formula Diagram, we also determine the formula of each intermediate variable and fill the Formula List (see Figure 8).

We found it preferable to have more variables with simple formulas than less variables with complex formulas. The golden rule of this methodology is to avoid having more than one kind of operator or function in a formula. Figure 9 shows the Formula Diagram of a non-recommended extreme case and Figure 10 its corresponding Formula List.

As the model grows in size, we can diagram sub-models separately and use connectors to indicate where intermediate variables are located. For example, Figure 11 shows a sub-model where the Total Distance is the result of a calculation instead of being an input variable. The Formula Diagram of Figure 7 can be modified as shown in Figure 12.

| Variable | Type | Formula / initial value |
|---|---|---|
| Daily Rate | Input | 58$ |
| Nb Days | Input, Interface | 12 |
| Daily Allowance | Input | 100 |
| Total Distance | Input, Interface | 1452 |
| Distance Cost | Input | 0,36$ |
| Rental Cost | Intermediate, Interface | = Nb Days * Daily Rate + IF(Total Distance > Nb Days * Daily Allowance, (Total Distance – Nb Days * Daily Allowance) * Distance Cost, 0) |

Figure 10 Formula List of the extreme case

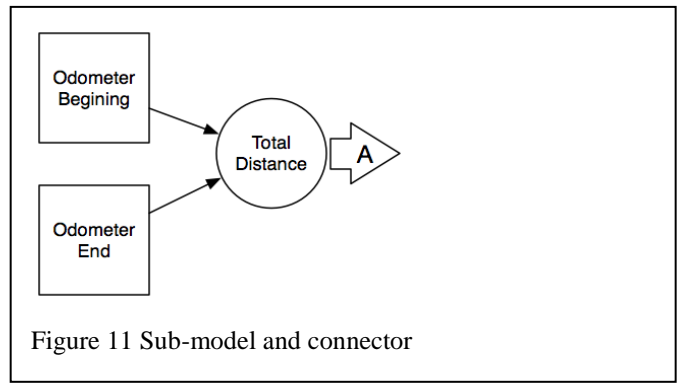

Figure 11 Sub-model and connector

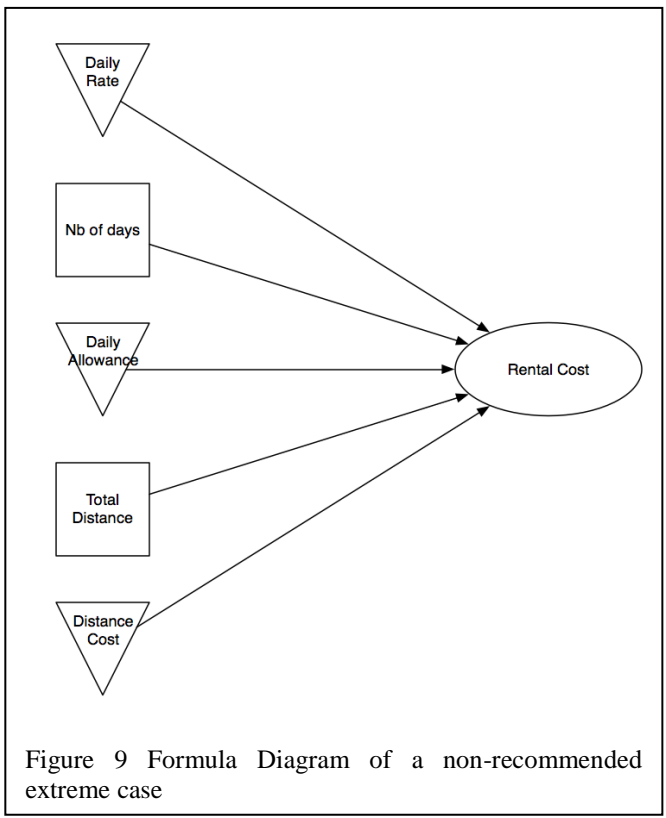

Figure 9 Formula Diagram of a non-recommended extreme case

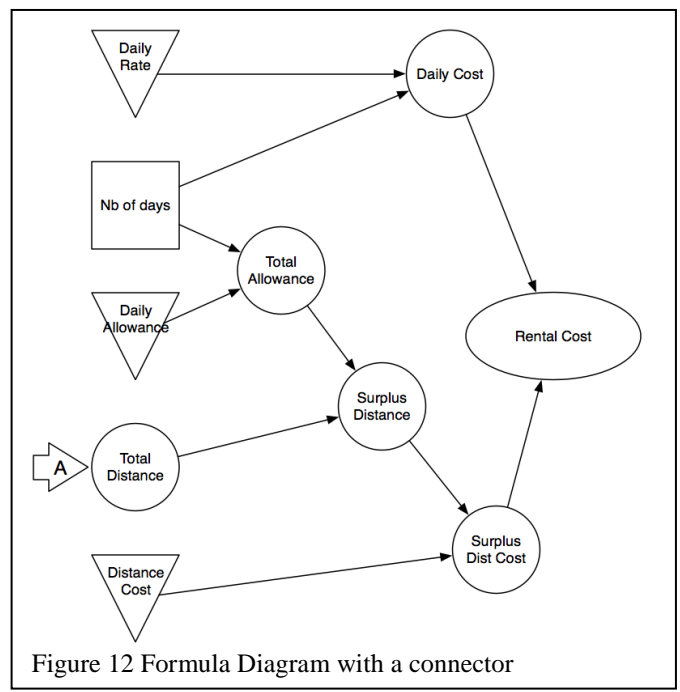

Figure 12 Formula Diagram with a connector

B. Implementation

1) Three-Tier Architecture

The Three-Tier Architecture was developed in the 1990's to improve the performance and the management of information systems. It consists of separating three major functions performed by systems and then building them separately, with the proper connections between them to ensure they work properly. These major functions are the *interface*, where the user interacts with the system, the *application*, which consists of the programs implementing the business logic required by the system and the *services*, which perform auxiliary services for the application, such as getting data to and from a database or accessing network resources.

The implementation of the Three-Tier Architecture in an Information System consists of developing separate computer programs. In our case, we will implement the tiers in separate sheets of the Excel file.

When you create a new file, Excel provides a sheet named Sheet1 and when you create new ones they are named Sheet2, Sheet3 and so on. You can rename a sheet simply by double-clicking on its name and typing a new name. To implement the three-tier architecture, you should at least have the following sheets: Interface, Model and Parameters as illustrated Figure 13. There may be some situations where you will use more than three sheets. For example, if you get data from different sources you may have multiple Parameters sheets in order for it to be easier to refresh their content at different moments. Also, if you have a complex model you may have multiple Model sheets to represent easier to manage sub-models. You would then need to name the sheets appropriately.

The Model sheet will have all the definition formulas of the intermediate variables (circles and ellipse in the Formula Diagram), presented in a specific block structure that is described in the next section. The Parameters sheets will simply have the input variables represented as triangles in the Formula Diagram. Finally, the Interface sheet will have the other input variables, represented as rectangles, and references to the intermediate variables the user wants to see, represented as ellipses in the Formula Diagram.

2) Model Sheet

Computer science has developed structured programming techniques in an effort to reduce the possibility of making errors. Using modules is one particular technique that consists of building a self-contained block in which all inputs are passed by value and the block's calculation does not use values that are not in its input list. The block is said to use local variables but there are exceptions where a global variable can be used, normally under controlled circumstances. A module that returns a value is called a function.

In the SSMI methodology, the implementation of a formula follows the structured programming technique of modules. It

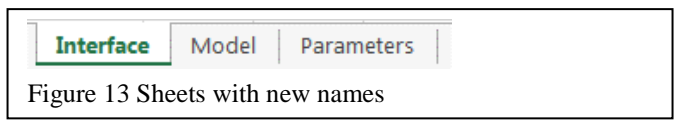
Figure 13 Sheets with new names

Figure 14 Block structure of the Model sheet

uses a block of rows with a precise structure, as illustrated in Figure 14. The bottom part of the block is the *definition part* and contains the variable that is being defined with a formula. In column A, we write a text label that will become the variable's name. In column B we write the formula that defines it, using only the cells in the top portion of the block.

The top part of the block is the *reference part* and contains all the variables used in the calculation of the variable we are defining. Each variable is identified by its name in column A and a simple reference formula in column B. The reference formula is simply a reference to the name of the cell where that variable is defined. Figure 15 displays the formula view and illustrates the reference formulas and the definition formula of a block.

Figure 15 Formula view of the block structure

Figure 16 Creating variable names

Figure 17 Visual clues to help audit a formula

Since all the definition formulas refer to cells immediately above, the formulas are easy to validate. They have a very low score according to the complexity measure developed by Hermans, Pinzger and van Deursen [8]

The reference and definition parts are visually separated by a line, easily implemented as a *top border* of the cells in the definition part. We format the rows with the definition formulas in bold. Furthermore, in Excel we name the cell containing the definition formula in column B with the text label of column A (see Figure 16). McKeever and McDaid [9] concluded that "range names do not improve the quality of spreadsheet developed by novice and intermediate users." Nevertheless we expect that our rules for naming cells and using those names only in the reference part of the blocks will be beneficial.

The block structure is similar to a function module in most programming languages. All the values that are passed to the function as arguments are in the reference part of the block and become the equivalent of local variables, and the function itself is implemented in the definition part of the block. Since all the cell references in the definition formula are in the immediate vicinity (see Figure 17) auditing a formula is easy.

3) Parameters Sheet

In the Parameters sheet, we write the name of each parameter in column A and put the respective value in column B, as illustrated in Figure 18. All the parameters are bold to indicate that it is here where they are defined and those representing monetary values should be formatted with the Currency style, where the number of decimals can be adjusted according to the value they represent.

All the parameters need to be named the same way

Figure 18 Parameters Sheet

Figure 19 Interface Sheet

Figure 20 Interface Sheet, formula view

intermediate variables are (see Figure 16).

4) Interface Sheet

In the Interface sheet, we write the name of each input variable in column A and put starting values in column B, as illustrated in Figure 19. The exact values are not important at this moment because the user will type proper values when he will use the spreadsheet. We will format in Bold the cells where each of our model's variables is defined. Finally, we format the cell with a proper number format, like the Currency or Percentage format.

When the model is completed, we return to the Interface sheet. Now, we can finish by putting a *reference formula* next to each output variable, referencing the cell where it is defined in the Model sheet, as demonstrated in Figure 20.

When the names of the input variables have been created, the Interface sheet can be redesigned to suit the user's preferences and needs: it doesn't need to be as austere as shown in these figures.

IV. Conclusion

In this paper, we have shown that we can apply principles and techniques from the fields of IS and SE to spreadsheet development. The SSMI methodology is concerned solely with the model; at the moment it does not address other issues like interface design and data import.

Further research should be done on the methodology's efficiency:

- Does it reduce the probability of making errors (logical and mechanical errors)?
- Is it easier to do maintenance on spreadsheets developed with it?
- Is it easier to detect errors?